\documentclass[%
reprint,
superscriptaddress,
showpacs,
 amsmath,amssymb,
 aps,
 pra,
floatfix,
a4paper,
twocolumn
]{revtex4-1}

\usepackage{graphicx}% Include figure files
\usepackage{dcolumn}% Align table columns on decimal point
\usepackage{bm}% bold math
\begin{document}

%\preprint{APS/123-QED}

%\textheight 24cm
%\textwidth 17cm
%\oddsidemargin 0cm
%\evensidemargin -1cm
%\topmargin -1cm
%opening
%\title{Single-shot shaped composite pulse for ultrafast high-fidelity robust quantum
%control}
%\title{Shaped pulse for ultrafast high-fidelity robust quantum control}
%
\title{Robust quantum control by shaped pulse}
\author{D. Daems}
\affiliation{Laboratoire Interdisciplinaire Carnot de Bourgogne, CNRS UMR 6303, Universit\'e de Bourgogne,
BP 47870, 21078 Dijon, France}
\author{A. Ruschhaupt}
\affiliation{Department of Physics, University College Cork, Cork, Ireland}
\author{D. Sugny}
\affiliation{Laboratoire Interdisciplinaire Carnot de Bourgogne, CNRS UMR 6303, Universit\'e de Bourgogne,
BP 47870, 21078 Dijon, France}
\author{S. Gu\'erin}
\email{sguerin@u-bourgogne.fr}
\affiliation{Laboratoire Interdisciplinaire Carnot de Bourgogne, CNRS UMR 6303, Universit\'e de Bourgogne,
BP 47870, 21078 Dijon, France}

%\date{}

\begin{abstract}
Considering the problem of the control of a two-state quantum system by an external field,
we establish a general and versatile method that allows the derivation of smooth pulses,
suitable for ultrafast applications, that feature the properties of high-fidelity, robustness,
and low area. Such shaped pulses can be viewed as a single-shot generalization of the composite
pulse-sequence technique with a time-dependent phase.
%For given efficiency and robustness, the obtained pulse remarkably outperform the
%known ones in term of areas.
\end{abstract}
\pacs{03.65.Aa, %Quantum systems with finite Hilbert space
32.80.Qk, %Coherent control of atomic interactions with photons
42.50.Dv %Quantum state engineering and measurements
42.50.Ex %Optical implementations of quantum information processing and transfer
} \maketitle

Modern applications of quantum control, such as quantum information processing \cite{NC},
require time-dependent schemes featuring three important issues: The transfer to the target
state should be achieved (i) with a high fidelity, typically with an admissible error lower than
$10^{-4}$ \cite{NC}, (ii) in a robust way with respect to the imperfect knowledge of the system or
to variations in experimental parameters, and (iii) with a minimum time of interaction and low field
energy in order to prevent unwanted destructive intensity effects.

The Rabi method (see e.g. \cite{Eberly}), corresponding to an exact resonant coupling between two
quantum states, leads to a complete transfer to an excited state from a ground state by a Rabi frequency
of area $A=\pi$ ($\pi$-pulse technique). This defines the transfer quantum speed limit in the sense that
the $\pi$-area of the Rabi frequency is the smallest area that gives a complete transfer $P=1$
\cite{Boscain}. Any (time-dependent) variation in the detuning requires a Rabi frequency area larger
than $\pi$, i.e. an increase of the fluence of the field, in order to recover the complete transfer.
This extra energy can be used to satisfy additional constraints such as robustness with respect to the
variations of parameters. In this framework, adiabatic techniques are famous examples \cite{adiab};
they however require in principle a large pulse area and do not lead to an exact transfer. Improvements
to optimize its efficiency by parallel adiabatic passage \cite{parallel} or by shortcuts to adiabaticity
\cite{shortcut}
have been proposed. Optimized shaping of the detuning and of the field amplitude leading to both robust
and precise transfer remains however an open question.

Making the transfer robust means that the errors that can appear in control parameters have to be
compensated, without even knowing these errors. A practical measure of the robustness can be defined
through the deviation of the excitation profile as a function of the considered parameters.
The use of composite pulses \cite{NMR} is a popular method for self-compensation of errors. The basic
idea consists in replacing the single resonant pulse driving the quantum transition by a sequence of
pulses with well-defined static phases. These phases, as well as the area of the individual pulses,
can be determined so that the derivatives of the excitation profile are nullified order by order by
increase of the number of pulses. Recently, an SU(2) algebraic approach has been extended for the design
of composite sequences of pulses with smooth temporal shapes and constant \cite{Torosov} or time-dependent
\cite{CAP} detuning. We emphasize that such smooth pulses are more appropriate for ultrafast processes,
for instance, with femtosecond pulses, requiring their production in the spectral domain \cite{femtotechno}, since rectangular pulses would lead to a prohibitively large spectrum. Typically, for a resonant complete inversion within a two-state system, while the profile $P$ deviates around $A=\pi$ as $P\sim 1- (A-\pi)^2$ with respect to the area for the Rabi method, it deviates around $A=\pi$ as $P\sim 1-(A-\pi)^{2n}$ with $n$ the (odd) number of composite resonant $\pi-$pulses.
Techniques of optimal control are also actively developed for this purpose of robustness \cite{NMR2}, but
they only lead to some purely numerical solutions without any insight into the physical mechanism nor the
analytical derivation of the control fields.

In this Letter, we establish a control strategy that allows a robust transfer by a pulse specifically
shaped in phase and amplitude. This can be viewed as a generalization of the strategy relying
on a series of composite pulses with static phases to a \emph{single-shot} pulse of
\emph{time-dependent phase}.
We first show that the issue of robustness can be reduced to nullifying the derivatives of the excitation
profile to a target state order by order. The central result of this work is that this can be in general achieved by an oscillatory parameterization of the phase of the wave function.
This continuous trigonometric basis is the key difference with respect to known methods which make use of stepwise functions (e.g. for the phase in the case of composite pulses).
This allows an explicit derivation of the components of the shaped pulse.
Our technique is explicitly shown for the robust complete inversion with respect to the pulse area, to the detuning, or to both parameters. This approach is however versatile and can be be applied to other types of robustness and to the transfer to more complicated targets, such as fully robust quantum gates in view of applications in quantum information processing.

The resulting smooth shaped pulses feature the required properties of high-fidelity, high-order robustness, and low area, in particular smaller than their composite counterpart for the same efficiency. Furthermore the pulses we derive have an explicit analytic form with very few parameters to adjust contrary to numerical optimal control procedures where a large number of parameters is used (see for instance \cite{Glaser}), and the resulting solution features a much simpler form than the ones usually obtained \cite{skinner2003,rabitz}.

Our discussion is based on a resonant system (rotating wave approximation)
between two states $|1\rangle$ and $|2\rangle$, for which the most general
Hamiltonian governing the dynamics can be written
\cite{Shore}
 \begin{equation}
\hat H(\Delta,\Omega,\eta)=\frac{\hbar}{2}\left[\begin{array}{cc} -\Delta(t) & \Omega(t)e^{-i\eta(t)}\\ \Omega(t)e^{i\eta(t)} & \Delta(t)\end{array}\right].
 \end{equation}
The Rabi frequency is decomposed into an absolute value $\Omega(t)>0$ (proportional to the field amplitude
for a one-photon transition), of area $A=\int\Omega(t) dt$, and a phase $\eta(t)$.
%This phase can also be moved to the
%detuning part by an appropriate transformation (see below).
The detuning $\Delta(t)$ between the field frequency and the transition features a time-independent
(static) detuning and a time-dependent part, produced, at the femtosecond timescale, by a
spectral phase shaping through the phase $\eta$ \cite{femtotechno}.

The errors arise from the imperfect knowledge of the area $A$ (for instance through an imperfect knowledge of the coupling constant), of the static detuning $\delta$ (often referring to a problem with an inhomogeneous broadening of an ensemble due to its environment), or to dynamical fluctuations of the pulse shape or of its instantaneous phase (corresponding to fluctuations of the time-dependent part of the detuning).
For simplicity, we focus our discussion on the deviation with respect to the area $A$ of the Rabi frequency and the static detuning $\delta$. Robust methods are designed to improve the quadratic deviation benchmark of the Rabi method.

The solution of the time dependent Schr\"odinger equation (TDSE)
$i\hbar\frac{\partial}{\partial t}\phi=H\phi$ can be
parameterized in the most general way with two angles
$\{\theta\equiv\theta(t)\in[0,\pi],\varphi-\eta\equiv
\varphi(t)-\eta(t)\in]-\pi,\pi]\}$
on the Bloch sphere of cartesian coordinates
(with $\rho_{mn}\equiv\langle m|\varphi\rangle\langle\varphi|n\rangle$)
%\begin{eqnarray}
$\rho_x=\rho_{21}+\rho_{12}=\sin\theta\cos\varphi$,
$\rho_y=i(\rho_{21}-\rho_{12})=\sin\theta\sin\varphi$,
$\rho_z=\rho_{11}-\rho_{22}=\cos\theta$,
%\end{eqnarray}
and with a global phase $\gamma\equiv\gamma(t)$ as
\begin{equation}
\label{solgen}
\phi=\left[\begin{array}{cc}e^{i\varphi/2}\cos(\theta/2)\\
e^{-i\varphi/2} \sin(\theta/2)\end{array}\right]e^{-i\gamma/2}.
\end{equation}
We first make a phase transformation
%\begin{equation}
%T=\left[\begin{array}{cc} e^{-i\alpha/2} & 0\\
%0 & e^{i\alpha/2}\end{array}\right],
%\end{equation}
$T=\hbox{diag}[e^{-i\eta/2}, e^{i\eta/2}]$
to deal with a real symmetric Hamiltonian
$H=T^{\dagger}\hat H T-i\hbar T^{\dagger}\frac{dT}{dt}
=\hat H(\Delta+\dot\eta,\Omega,0)$
%\begin{equation}
%H=T^{\dagger}\hat H T-i\hbar %T^{\dagger}\frac{dT}{dt}=\frac{\hbar}{2}\left[\begin{array}{cc}
%-(\Delta+\dot\eta) & \Omega\\
%\Omega & \Delta+\dot\eta\end{array}\right]
%=\hat H(\Delta+\dot\eta,\Omega,0),
%\end{equation}
of solution $\psi=T^{\dagger}\phi$.
%\begin{align}
%\label{sol}
%\psi=T^{\dagger}\phi&
%%=\left[\begin{array}{cc}e^{i\alpha/2}\cos(\theta/2)\\
%%e^{-i(\varphi+\alpha/2)} \sin(\theta/2)\end{array}\right]
%%e^{-i\int^t ds\lambda(s)},\\
%%&
%=\left[\begin{array}{cc}e^{i(\varphi+\eta)/2}\cos(\theta/2)\\
%e^{-i(\varphi+\eta)/2} \sin(\theta/2)\end{array}\right]
%e^{-i\gamma/2}.
%\end{align}
This shows that the phase $\eta$ can be incorporated in the detuning term and simply interpreted as the rotation of the axes $x$ and $y$ about the $z$-axis of the Bloch sphere.
%This means that one can control the relative phase between the two components by an
%appropriate choice of $\eta$, if we do not consider the control of the global phase.
%In this work, since we do not consider the robustness with respect to $\eta$, we %can assume that $\eta=0$.
One can thus assume $\eta=0$ without loss of generality.
Inserting this solution in the TDSE, we get:
%\begin{align}
%\frac{d}{dt}\theta&=\Omega\sin(\alpha+\varphi),\\
%\lambda&=\frac{\Omega}{2}\frac{\cos(\alpha+\varphi)}{\sin\theta}-\frac{1}{2}
%\frac{d}{dt}\varphi,\\
%\frac{d}{dt}(\alpha+\varphi)&=\Delta+\Omega\cos(\alpha+\varphi)\,\text{cotan}\,
%\theta
%\end{align}
%OR
\begin{subequations}
\label{eqtpl}
\begin{align}
\label{eqtheta}
\dot\theta&=\Omega\sin\varphi,\\
\label{eqphi}
\dot\varphi&=\Delta+\Omega\cos\varphi\,\text{cotan}\,\theta,\\
\label{eqlambda}
\dot\gamma&
=\Omega\frac{\cos\varphi}{\sin\theta}=\dot\theta\,\frac{\hbox{cotan}\,\varphi}{\sin\theta}.
%-\dot\varphi
%=\Omega\cos\varphi\,\tan(\theta/2)-\Delta.
\end{align}
\end{subequations}
%which equivalently leads for $\lambda$ to
%\begin{align}
%\lambda&
%=-\frac{1}{2}\Delta-\frac{\Omega}{2}\cos\varphi\,\text{cotan}\,(\theta/2).
%\end{align}
%ou (Yavor) ?
%\begin{align}
%\lambda&
%=-\frac{1}{2}\Delta-\frac{\Omega}{2}\cos\varphi\,\tan(\theta/2).
%\end{align}
We assume $\phi(t_i)=|1\rangle$ as initial condition, corresponding to the initial conditions
% \begin{equation}
$ \theta(t_i)\equiv\theta_i=0$ (north pole), $\varphi(t_i)\equiv\varphi_i=\gamma(t_i)\equiv\gamma_i$ (not specified by the initial state).
%\end{equation}
We consider the complete population transfer, which is achieved for the final condition $\theta(t_f)\equiv\theta_f=\pi$ (south pole).
For any trajectory featured by $\varphi(t)$, $0\leftarrow\theta(t)\rightarrow\pi$ on the Bloch sphere, one can
integrate \eqref{eqtheta}:
%\begin{equation}
$\theta_f-\theta_i=\pi=\int_{t_i}^{t_f}ds\,\Omega(s)\sin\varphi(s)$.
%\end{equation}
Since $\int_{t_i}^{t_f}ds\,\Omega(s)\sin\varphi(s)\le \int_{t_i}^{t_f}ds\,\Omega(s)$,
one concludes that a complete population transfer is achieved
when $\int_{t_i}^{t_f}ds\,\Omega(s)\ge\pi$, in consistency with Ref.
\cite{Boscain}. The minimum area $\pi$ is obtained for the
meridian $\varphi=\pi/2$, implying $\gamma_i=\pi/2$, $\dot\gamma=0$ from \eqref{eqlambda} and $\Delta=0$ from \eqref{eqphi}, which corresponds to the Rabi transfer.
On the other hand, the ideal adiabatic solution is derived when $A\gg1$ giving $\varphi\to0,\pi$ [in order to have a bounded $\dot\theta$ from \eqref{eqtheta}], that is $\gamma(t)\to\pm\int_{t_i}^{t}\sqrt{\Omega(s)^2+\Delta(s)^2}ds$ from \eqref{eqlambda}. From \eqref{solgen}, we recover the well-known adiabatic dynamics.

More generally, the transfer to a given target state up to a (global) phase corresponds to the desired final conditions $\theta_f$ and $\varphi_f$. The control of the phase of the target state would require the additional condition $\gamma(t_f)\equiv\gamma_f$, which will not be considered here. We will additionally assume $\Omega(t_i)=\Omega(t_f)=0$ in view of a practical implementation. We will also consider a monotonic increasing of $\theta$, i.e. $\dot\theta>0$, with $\Omega\ge0$, which leads to $0\le\varphi\le\pi$ from Eq. \eqref{eqtheta}.

\begin{figure}[!h]
\begin{center}
\includegraphics[scale=0.9]{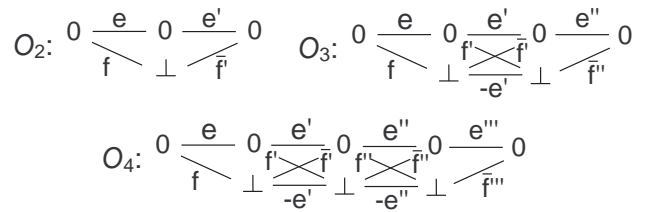}
\caption{Symbolic path-diagrams giving the construction of the $O_n$ integrals. The symbol $e$ stands for $e(t)$, $e'$ for $e(t')$, and so on. For instance, for $n=3$ ($n=4$), the diagram features 4 (8) paths. Its extension for larger $n$ is direct.}
\label{diag}
\end{center}
\end{figure}

The goal of the control consists in finding a particular solution of \eqref{eqtpl}
with the final  conditions corresponding to the target state, which is robust and
of lowest $\Omega$ area.
The strategy is similar to the one used for generating composite pulses: One nullifies the derivatives of the transfer profile with respect to the considered parameters. This can be here achieved with the Hamiltonian of the form
% \begin{equation}
%H_{\varepsilon,\delta}(t)=\frac{\hbar}{2}\left[\begin{array}{cc} -(\Delta(t)+\delta) & %\Omega(t)(1+\alpha)\\ \Omega(t)(1+\alpha) & \Delta(t)+\delta\end{array}\right].
%%+\varepsilon_{z}\left[\begin{array}{cc} -1 & 0\\ 0 &
%%1\end{array}\right]+\varepsilon_{x}\left[\begin{array}{cc} 0& \Omega\\ \Omega &
%%0\end{array}\right].
% \end{equation}
 \begin{equation}
 \label{Halpdelt}
H_{\alpha,\delta}(t)=\frac{\hbar}{2}\left[\begin{array}{cc} -\Delta(t) & \Omega(t)\\ \Omega(t) & \Delta(t)\end{array}\right]
+\frac{\hbar}{2}\left[\begin{array}{cc} -\delta & \alpha\Omega(t)\\ \alpha\Omega(t) & \delta\end{array}\right].
 \end{equation}
% \begin{equation}
%H(t)=\frac{\hbar}{2}\left[\begin{array}{cc} -\Delta(t) & \Omega(t)\\ \Omega(t) &
%\Delta(t)\end{array}\right]
%+\varepsilon V,\ \varepsilon V=\frac{\hbar}{2}\left[\begin{array}{cc} -\delta &
%\alpha\Omega(t)\\ \alpha\Omega(t) & \delta\end{array}\right].
% \end{equation}
The corresponding solution $\phi_{\alpha,\delta}(t)$ is parameterized by $\alpha$ and $\delta$ for given functions $\Omega(t)$ and $\Delta(t)$.
We assume that $\Omega(t)$ and $\Delta(t)$ are such that the complete transfer to the target state $\phi_T$ is achieved at $t=t_f$ for $\delta=0$ and $\alpha=0$: $\phi_{0}(t_f)=\phi_T$.
%One can make a Taylor expansion:
%%\begin{align}
%%\nonumber
%%%&\phi_{\varepsilon,\delta}(t_f)=\phi_{0,0}(t_f)+\varepsilon
%%%\left.\frac{\partial\phi_{\varepsilon,\delta}(t_f)}
%%%{\partial\varepsilon}\right|_{0,0}
%%%+\delta
%%%\left.\frac{\partial\phi_{\varepsilon,\delta}(t_f)}
%%%{\partial\delta}\right|_{0,0}\\
%%%&+\frac{1}{2}\left(\left.\frac{\partial^2\phi_{\varepsilon,\delta}(t_f)}
%%%{\partial\varepsilon^2}\right|_{0,0}
%%%+\left.\frac{\partial^2\phi_{\varepsilon,\delta}(t_f)}
%%%{\partial\delta^2}\right|_{0,0}+
%%%2\left.\frac{\partial^2\phi_{\varepsilon,\delta}(t_f)}
%%%{\partial\varepsilon\delta}\right|_{0,0}\right)
%%&\phi_{\varepsilon,\delta}(t_f)=\phi_T+\varepsilon
%%\frac{\partial\phi_{0,0}(t_f)}
%%{\partial\varepsilon}
%%+\delta
%%\frac{\partial\phi_{0,0}(t_f)}
%%{\partial\delta}\\
%%&+\frac{1}{2}\left(\frac{\partial^2\phi_{0,0}(t_f)}
%%{\partial\varepsilon^2}
%%+\frac{\partial^2\phi_{0,0}(t_f)}
%%{\partial\delta^2}+
%%2\frac{\partial^2\phi_{0,0}(t_f)}
%%{\partial\varepsilon\partial\delta}\right)+\cdots
%%\end{align}
%\begin{align}
%&\phi_{\varepsilon,\delta}(t_f)=\phi_T+\mathbf{r}^T\mathbf{\nabla}\phi_f
%+\frac{1}{2!}\mathbf{r}^T\mathbf{\nabla}^2\phi_f\mathbf{r}+\cdots
%\end{align}
%with $\mathbf{r}=[\alpha\ \delta]^T$ and the shorthand notation
%$\mathbf{\nabla}\phi_f\equiv
%\mathbf{\nabla}\phi_{\alpha,\delta}(t_f)|_{\varepsilon=0,\delta=0}$ and
%$\mathbf{\nabla}^2\phi_f$ the corresponding Hessian.
One makes a perturbative expansion of $\phi_{\alpha,\delta}(t_f)$ with respect to $\alpha$ and $\delta$ taking the second matrix term of \eqref{Halpdelt} as a perturbation denoted $V$ (the small parameters are both $\alpha$ and $\delta$ assumed of the same order) \cite{robust}:
%\begin{equation}
$\langle\phi_T|\phi_{\alpha,\delta}(t_f)\rangle=1+O_1+O_2+O_3+\cdots,$
%\end{equation}
where $O_n$ denotes the term of total order $n$, giving for the excitation profile
\begin{align}
\label{pertexp}
%\nonumber
|\langle\phi_T|\phi_{\alpha,\delta}(t_f)\rangle|^2=1+ \widetilde O_1+
\widetilde O_2+\widetilde O_3+\cdots
\end{align}
with $\widetilde O_n$ the term of order $n$. The first two terms read
\begin{subequations}
\begin{align}
&O_1=-i\int_{t_i}^{t_f}\langle\phi_0(t)|V(t)|\phi_0(t)\rangle dt\equiv
-i\int_{t_i}^{t_f}e(t)dt,\\
&O_2=(-i)^2\int_{t_i}^{t_f}dt\int_{t_i}^{t}dt'[e(t)e(t')
+f(t)\bar f(t')],
%\\
%&O_3=(-i)^3\int_{t_i}^{t_f}dt\int_{t_i}^{t}dt'\int_{t_i}^{t'}dt''[e(t)e(t')e(t'')
%\nonumber\\
%&+e(t)f(t')\bar f(t'')
%+f(t)\bar f(t')e(t'')-f(t)e(t')\bar f(t'')]
\end{align}
\end{subequations}
with $\widetilde O_1=O_1+\bar O_1$, $e=-\frac{1}{2}(\delta\cos\theta-\alpha\dot\gamma\sin^2\theta)$,
%\begin{equation}
%\label{deff}
%f=\langle\phi_0|V|\phi_{\perp}\rangle=\frac{1}{2}[\delta\sin\theta
%+\alpha\Omega(\cos\theta\cos\varphi-i\sin\varphi)]e^{i\gamma},
%\end{equation}
\begin{equation}
\label{deff}
f=\langle\phi_0|V|\phi_{\perp}\rangle=\frac{1}{2}\Bigl[\delta\sin\theta
+\alpha\Bigl(\frac{1}{2}\dot\gamma\sin2\theta-i\dot\theta\Bigr)\Bigr]e^{i\gamma},
\end{equation}
and the orthogonal solution of the TDSE $\phi_{\perp}(t)=\left[e^{i\varphi/2}\sin(\theta/2),-e^{-i\varphi/2} \cos(\theta/2)\right]^Te^{i\gamma/2}$ such that $\langle \phi_{\perp}(t)|\phi_{0}(t)\rangle=0$.
The other terms can be determined from the symbolic diagrams depicted in Fig. \ref{diag}.
Since $e(t)$ is real, there is no first-order deviation for the excitation profile.
One can simplify the second order as
\begin{equation}
\label{secondorder}
\widetilde O_2\equiv O_2+\bar O_2+\bar O_1O_1=-\Bigl|\int_{t_i}^{t_f}f(t)dt\Bigr|^2
\end{equation}
using the property $\int_{\tau}^{T}dt\int_{\tau}^{t}dt'[a(t)b(t')
+a(t')b(t)]=\int_{\tau}^{T} a(t)dt\int_{\tau}^{T} b(t)dt$. This property also implies
$\int_{\tau}^{T}dt\ a(t)\int_{\tau}^{T}dt\, b(t)\int_{\tau}^{t}dt'c(t')=\int_{\tau}^{T}dt\int_{\tau}^{t}dt'\int_{\tau}^{t'}dt''[a(t)b(t')c(t'')
+b(t)a(t')c(t'')+b(t)c(t')a(t'')]$ and extends as
$\int_{\tau}^{T}dt\int_{\tau}^{t}dt'\int_{\tau}^{t'}dt''
\sum_{\sigma} a(t^{(\sigma(0))})b(t^{(\sigma(1))})c(t^{(\sigma(2))})
=\int_{\tau}^{T} a(t)dt\int_{\tau}^{T} b(t)dt\int_{\tau}^{T} c(t)dt$,
where $\sum_{\sigma}$ means the summation over the six permutations of the number of primes
(between 0 and 2).
This is used to determine relatively simple integrals for higher orders. The third order reads:
\begin{equation}
\label{thirdorder}
\widetilde O_3=-4\int_{t_i}^{t_f}dt\int_{t_i}^{t}dt'\int_{t_i}^{t'}dt'' \text{Im}[\bar f(t) e(t') f(t'')].
\end{equation}
Robustness at a given order $n$ is obtained when the parameters of the field are chosen such that they
allow nullifying the integrals $\widetilde O_m$ for $m\le n$. The second-order robustness
issue corresponds to the two equations:
%\begin{subequations}
%\label{eqf}
%\begin{eqnarray}
%\int_{t_i}^{t_f}e^{i\gamma}\sin\theta\, dt&=&0,\\
%4\int_{\theta_i}^{\theta_f}
%e^{i\widetilde\gamma}\sin^2\theta\, d\theta&=&[\sin2\theta]_{\theta_f}^{\theta_i}
%\end{eqnarray}
%\end{subequations}
\begin{equation}
\label{eqf}
\int_{t_i}^{t_f}e^{i\gamma}\sin\theta\, dt=0,\
\int_{\theta_i}^{\theta_f}
e^{i\widetilde\gamma}\sin^2\theta\, d\theta=\frac{1}{4}
[e^{i\widetilde\gamma}\sin2\theta]_{\theta_f}^{\theta_i}
\end{equation}
for the robustness with respect to the detuning $\delta$, and to the pulse area respectively.
If one considers the robustness only with respect to the pulse area, the corresponding equation
involves an integral which is independent of the particular temporal parameterization of $\theta$.
For some integrals, such as, for instance the second order robustness with respect to the detuning
[left equation of \eqref{eqf}], we need additionally an explicit time parameterization of $\theta$.
For the Rabi method ($\Delta=0$), these equations \eqref{eqf} cannot be satisfied for any pulse shape:
It is robust up to the first order for the excitation profile.
Below, we explicitely derive solutions
for the issue of inversion (with $\theta$ varying from 0 to $\pi$). The same technique applies for other
target states.

%Similar formulas give the fourth order
%\begin{align}
%\label{fourthorder}
%&\int_{t_i}^{t_f}dt\int_{t_i}^{t}dt'\int_{t_i}^{t'}dt''\int_{t_i}^{t''}dt''' %\{2\,\text{Re}[f(t)\bar f(t') f(t'') \bar f(t''')]\nonumber\\
%&+8\,\text{Re}[f(t)e(t') e(t'') \bar %f(t''')]\}+\Bigl|\int_{t_i}^{t_f}dt\int_{t_i}^{t}dt' f(t) \bar f(t')\Bigr|^2.
%\end{align}
%\begin{equation}
%\delta\int_{t_i}^{t_f}e^{i\gamma}\sin\theta\,dt-i\alpha\Bigl(2\int_{\theta_i}^{\theta_f}
%e^{i\widetilde\gamma}\sin^2\theta\ d\theta+\frac{1}{2}\sin2\theta_f
%e^{i\gamma_f}\Bigr)=0
%\end{equation}

These equations \eqref{eqf} (and the ones corresponding to higher orders) can be exactly solved by an
oscillatory parameterization with a Fourier series of the global phase as a function of $\theta$,
$\gamma(t)\equiv\widetilde\gamma(\theta)$ (with the label $a$ or
$b$ to identify it):
\begin{subequations}
\label{gamtilde}
\begin{align}
\label{gamtildea}
\widetilde\gamma_a(\theta)&=2\theta+C_1\sin(2\theta)+\cdots
+C_n\sin(2n\theta)+\cdots\\
\label{gamtildeb}
\widetilde\gamma_b(\theta)&=\theta+C_1\sin(2\theta)+\cdots
+C_n\sin(2n\theta)+\cdots
\end{align}
\end{subequations}
choosing the smooth parameterization $\theta(t)=\pi(\,\hbox{erf}(t/T)+1)/2$ (which will lead to a
smooth pulse) with $T$ featuring the duration of interaction.  These parameterizations \eqref{gamtilde}
allow a simple expression in view of controlling the phases, since, from \eqref{eqlambda}, we have
$\text{cotan}\,\varphi=\sin(\theta)d\widetilde\gamma/d\theta$.
These choices give
$\gamma_f=2\pi$ ($\pi$) for parameterization \eqref{gamtildea}
[\eqref{gamtildeb}]. For the choice \eqref{gamtildea}, the imaginary parts of Eq. \eqref{eqf} are
nullified for all $C_n$'s. On the other hand, for \eqref{gamtildeb}, the real parts are nullified
for all $C_n$'s.
In both cases, one has then to adjust the coefficients $C_{n}$ to nullify the other integrals.
An important result is that, due to the oscillatory nature of the parameterization, only a few
coefficients are needed to nullify the integrals; typically one additional coefficient different
from zero is considered to nullify another single (real) integral.

\begin{table}[h]
\caption{Robustness of order $n$ (i.e. $\widetilde O_{m\le n}=0$)
with respect to the area (referred to as type $A$),
to the detuning (type $\delta$), or both (type $A\delta$), with the parameterization
\eqref{gamtildea} (referred to as param. $a$) or \eqref{gamtildeb} (param. $b$), and the coefficients $C_j$, $j=1,2,3$, ($C_{j>3}=0$). We have considered parameterizations leading to low pulse areas.} %title of the table
\centering % centering table
\begin{tabular}{ccccccc} % creating eight columns
\hline\hline %inserting double-line
 Type & Param. & Order &
%\multicolumn{7}{c}{Sum of Extracted Bits} \\[0.5ex]
    $C_1$ & $C_2$ & $C_3$
    & Pulse area ($\times\pi$) \\ [0.5ex]
\hline % inserts single-line
$A$ & a & 3 &-1 & 0& 0& 2.16 \\ % Entering row contents
$A$ & b & 3 &-1.6788 & 0& 0& 2.09 \\
$\delta$ & a & 3 &-0.2305 & 0& 0& 1.78 \\
$A\delta$ & b & 2 &-1.189 & 0.7285 & 0 & 2.23\\
$A$ & a & 5 &-2.4864 & -0.74 & 0& 3.14 \\
$A$ & a & 7 &-3.46 & -1.365 & -0.5 & 3.86 \\ [1ex] % [1ex] adds vertical space
\hline % inserts single-line
\end{tabular}
\label{Tresult}
\end{table}

Some obtained coefficients are summarized in Table \ref{Tresult}. The resulting pulse area
$\int_{t_i}^{t_f}\Omega(t) dt=\int_0^{\pi}\sqrt{1+(\dot{\widetilde\gamma})^2\sin^2\theta }\,d\theta$
is mentioned. One can notice increasing areas when more coefficients different from zero are considered.
Table \ref{Tresult} shows for instance that one obtains a pulse area of only $1.78\pi$ for the
\emph{third order robustness} [corresponding to the integrals \eqref{secondorder} and \eqref{thirdorder}
nullified, i.e. up to the third order] solely with respect to the detuning taking the parameterization
\eqref{gamtildea} with $C_1=-0.2305$. Alternatively, we obtain at best a pulse area of $2.09\pi$ for the
third order robustness solely with respect to the pulse area taking the parameterization \eqref{gamtildeb}.
We emphasize that this latter choice of parameterization allows a smooth pulse of slightly smaller area
than the non-smooth pulse proposed in \cite{robust} (with $C_1=-1$, see first line of Table \ref{Tresult}).
In both cases, the third order integrals \eqref{thirdorder} are
systematically nullified when the coefficients are adjusted to nullify the second order
\eqref{secondorder} due to the symmetry of the parameterization.
One can alternatively force robustness with respect to both the detuning $\delta$ and the pulse area,
nullifying both terms of Eq. \eqref{eqf} (see fourth line of Table \ref{Tresult}). This shows the
remarkable result that the obtained pulse area is only slightly larger than the one obtained above for
robustness solely with respect to the pulse area (but for a robustness of order 2).

%\begin{equation}
%f=\frac{1}{2}[\delta\sin\theta+\alpha\Omega(\cos\theta\cos\varphi-i\sin\varphi)]e^{i\gamma}
%\end{equation}

If one considers robustness at high order, one has to nullify the higher order terms.
%[see Eq. \eqref{fourthorder} for the fourth order].
The results for the robustness with respect to the pulse area at orders 5 and 7 are shown in Table \ref{Tresult}.

\begin{figure}[!h]
\begin{center}
\includegraphics[scale=0.65]{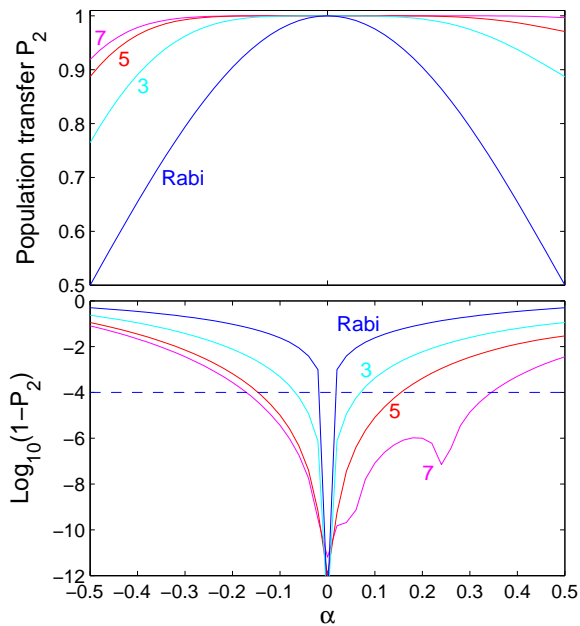}
\caption{Robustness with respect to the pulse area: Population transfer $P_2$ at the end of the pulse as a function of the relative deviation $\alpha$ (dimensionless) of the area. Different orders of robustness (from order 3 to 7, as indicated next to the curves), where the used coefficients and parameterizations are indicated in Table \ref{Tresult}, are compared to the standard Rabi method. Upper: Linear scale of $P_2$. Lower: Logarithm to the basis 10 of the deviation.}
\label{Figrobust}
\end{center}
\end{figure}

The robustness with respect to pulse area at different orders is demonstrated in Fig. \ref{Figrobust}. It shows the numerical deviation of the transfer profile for various values of $\alpha$ as defined in Eq. \eqref{Halpdelt} (here we take $\delta=0$), with the parameters of Table \ref{Tresult} and the parameterization \eqref{gamtildea}. As expected, the transfer profile becomes flatter and flatter when one nullifies higher order terms in the perturbative expansion \eqref{pertexp} (see upper frame of Fig. \ref{Figrobust}). We can notice that the profile is not symmetric with respect to $\alpha=0$. More precisely, the transfer is less robust for a smaller area (corresponding to negative values of $\alpha$). This result is in accordance with the expected property that the robustness is better for a larger pulse area. The lower frame Fig. \ref{Figrobust} depicts the deviation of the excitation profile at a logarithmic scale. It shows the remarkable result that the population transfer is accomplished with the $10^{-4}$ high-fidelity accuracy benchmark (indicated by the horizontal dashed line in Fig. \ref{Figrobust}) even with an error in area up to $17\%$ for the highest order solution (limited by a negative area deviation).

\begin{figure}[!h]
\begin{center}
\includegraphics[scale=0.65]{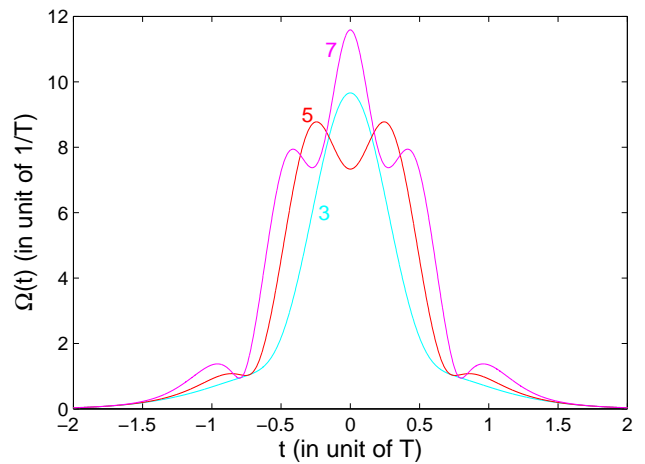}
\caption{Derived Rabi frequency pulse shapes for robustness of orders 3, 5 and 7 with respect to the pulse area (with the parameters of Table \ref{Tresult}).}
\label{Figshape}
\end{center}
\end{figure}

Figure \ref{Figshape} displays the shapes of the Rabi frequencies that are derived:
They oscillate more and more with a larger area for a higher order robustness.
These oscillating pulses can be interpreted as an efficient generalization of the
composite pulse sequence, but with time-dependent phases instead of the static ones
used for the composite pulses.

In conclusion, we have constructed and analyzed a generalization of the technique of
composite pulse sequence with static phases to a single-shot pulse of time-dependent phase.
We have reduced the robustness issue to a problem of nullifying integrals, and have shown
that this can be achieved for an oscillatory parameterization of the phase of the solution.
The resulting smooth shaped pulse, suitable for
ultrafast applications, feature properties of high-fidelity, high-order robustness, and low area.
Furthermore the pulses we derive have an explicit and relatively simple form which could
be easily implemented experimentally. A complete analysis has been given for the robustness
with respect to the pulse area, to the detuning, or to both parameters, and an explicit parameterization
has been demonstrated for the inversion problem. Here only two or three free parameters $C_n$
have been systematically adjusted to achieve a very efficient robust inversion. The techniques
which we present is however versatile and can be applied to other types of robustness and target,
such as coherent superposition of states or simultaneous control.
Using the same parameterization, one can for instance derive a shaped pulse, with a few coefficients
to be adjusted, producing a robust superposition of state [with $\theta$ varying from 0 to a given value
featuring the weight of the desired superposition, and $\varphi_f$ chosen from \eqref{eqlambda}].
Ultimately, using the same technique (but adapting the parameterization), one can also produce robust
propagators with self-compensation of errors, suitable for quantum information processing.
In this case, more integrals to be nullified will be involved (but of the same type that the ones
described in this work), which will need more coefficients $C_n$ to be adjusted: algorithms of optimal
control will be then needed to optimize these coefficients \cite{skinner2010}.
This task will however be relatively easy in comparison with traditional optimal
control techniques which need in general more than hundred coefficients to be
adjusted. This simplification will be also useful in view of extension
to more complicated systems.

\subsection*{Acknowledgments}

We acknowledge support from the European Marie Curie Initial Training Network GA-ITN-214962-FASTQUAST, and from the Conseil R\'egional de Bourgogne.

\end{document}